\documentstyle[psfig]{l-aa}
\newcommand{\be}{\begin{equation}}
  \newcommand{\en}{\end{equation}}
\thesaurus{13(13.25.5, 08.14.1, 08.09.2 1E1743.1--2843)}

\begin{document}
\def\ltsima{$\; \buildrel < \over \sim \;$}
\def\lsim{\lower.5ex\hbox{\ltsima}}
\def\gtsima{$\; \buildrel > \over \sim \;$}
\def\gsim{\lower.5ex\hbox{\gtsima}}
\def\spose#1{\hbox to 0pt{#1\hss}}
\def\approxlt{\mathrel{\spose{\lower 3pt\hbox{$\sim$}}
    \raise 2.0pt\hbox{$<$}}}
\def\approxgt{\mathrel{\spose{\lower 3pt\hbox{$\sim$}}
    \raise 2.0pt\hbox{$>$}}}
\def\deg {^\circ}
\def\mdot {\dot M}
\def\kms {$\sim$km$\sim$s$^{-1}$}
\def\gs {$\sim$g$\sim$s$^{-1}$}
\def\ergs {$\sim$erg$\sim$s$^{-1}$}
\def\cmtre {$\sim$cm$^{-3}$}\def\nupa{\vfill\eject\noindent}
\def\der#1#2{{d #1 \over d #2}}
\def\l#1{\lambda_{#1}}
\def\grb{$\gamma$-ray burst}
\def\grbs{$\gamma$-ray bursts}
\def\rosat{{\sl ROSAT} }
\def\cmdue {cm$^{-2}$}
\def\gcm {$\sim$g$\sim$cm$^{-3}$}
\def\rsole{$\sim$R_{\odot}}
\def\msole{$\sim$M_{\odot}}
\def\aa #1 #2 {A\&A, {#1}, #2}
\def\mon #1 #2 {MNRAS, {#1}, #2}
\def\apj #1 #2 {ApJ, {#1}, #2}
\def\aj #1 #2 {AJ, {#1}, #2}
\def\nat #1 #2 {Nature, {#1}, #2}
\def\pasj #1 #2 {PASJ, {#1}, #2}
\newfont{\mc}{cmcsc10 scaled\magstep2}
\newfont{\cmc}{cmcsc10 scaled\magstep1}
\newcommand{\bc}{\begin{center}}
  \newcommand{\ec}{\end{center}}

\title{$BeppoSAX$ and $ASCA$ observations of the Galactic Bulge source 1E~1743.1-2843}

\author{D.I. Cremonesi\inst{1}, S. Mereghetti\inst{2}, L. Sidoli\inst{2,3} and
 G.L. Israel\inst{4}.   }

\institute{
  {INTEGRAL Science Data Centre, Chemin d'Ecogia 16, Versoix, CH-1290,
    Switzerland}\and
  {Istituto di Fisica Cosmica ``G.Occhialini'', CNR, Via Bassini 15, I-20133 Milano,
    Italy}\and
  {Dipartimento di Fisica, Universit\`a di Milano, Via Celoria 16, I-20133
    Milano, Italy} \and
  {Osservatorio Astronomico di Roma, via dell'Osservatorio 2, I-00040 Monteporzio
    Catone (Roma), Italy}
  }

\maketitle
\label{sampout}
\begin{abstract}

  We report the results of $BeppoSAX$ and $ASCA$ observations  of the   X--ray source 1E~1743.1--2843,  located  in the
  region of the galactic center, performed between 1993 and 1998.
  Observations spanning almost twenty years show that 1E~1743.1--2843 is
  a persistent X--ray source, most likely  powered by accretion
  onto a compact galactic object.
  The absence of periodic pulsations and the relatively soft X--ray spectrum
  favour a Low Mass X--ray Binary containing a neutron star.
  However, no Type~I X--ray bursts have been observed in more than 650
  ks of observations, leaving other interpretations possible.
 
  \keywords{X--Rays -- Individual: 1E1743.1--2843}
  
\end{abstract}

\section{Introduction}

The source 1E~1743.1--2843 was discovered during the first soft
X--ray  imaging observations of the galactic center region
obtained with the Einstein Observatory (Watson et al. 1981).
Its  column density ($\gsim  10^{23}$ cm$^{-2}$) is one
of the highest observed in the
X--ray sources of this region of sky, suggesting a distance
similar or even greater than the galactic center.
To power the observed luminosity $\gsim$ 10$^{36}\times~d_{10\rm{kpc}}^{2}$ erg s$^{-1}$ 
accretion onto a compact object is clearly required.
Contrary to many other
X--ray binaries in the galactic center region that are characterized by
transient activity, 1E~1743.1--2843 is a persistent source, that has been detected
in all the observations carried out up to now
(Watson et al. 1981, \cite{1987Natur.330..544S},
Kawai et al. 1988, Sunyaev et al. 1991, Pavlinsky et al. 1994, Lu et al. 1996).

We have recently obtained a long pointing centered on  1E~1743.1--2843
in the context of our survey of the galactic center region with the
$BeppoSAX$
satellite (Sidoli et al. 1998a). In addition, several other
$BeppoSAX$ and $ASCA$
serendipitous
observations of this source are available, spanning the years from 1993 to
1998.

\section{Observations and Data Analysis}

In Table 1 we summarize the $BeppoSAX$ and
$ASCA$ observations of 1E~1743.1--2843 used
in this work.

In addition to the 70.4 ks long pointing obtained on April 13-15, 1998,
for which 1E~1743.1--2843 was the main target (S4),
four other $BeppoSAX$ observations
are available: one pointed on the galactic center
(SgrA West, $\sim$100 ks) and
three shorter ones pointed at different locations along the galactic
plane in order to map the  Sgr B2 molecular cloud.\\
Our analysis of the  $BeppoSAX$ observations is based on data obtained with
the MECS instrument (Boella et al. 1997).
The MECS instrument consists of three  position-sensitive
gas-scintillation proportional counters providing images in the  
1.3--10 keV energy range within a field of view of 56' diameter,
and with an angular resolution of $\gsim$1' (50\% power radius of 
 $\sim$75'' at 6 keV, on-axis).
The MECS energy resolution is $\sim$8.5$\sqrt{6/{\rm E_{keV}}}$\%
FWHM.
After the failure of MECS1 in May 1997, only two units are available.

The $ASCA$ satellite (Tanaka, Inoue \& Holt 1994)
observed  the region of sky
containing 1E~1743.1--2843 several times.
$ASCA$    provides simultaneous data in four
coaligned telescopes, equipped with two solid state detectors
(SIS) and two gas scintillation proportional counters
(GIS).
However, since in most of these observations 1E~1743.1--2843 lies at large
off-axis angles, only
the GIS instruments, with a larger field of view
(44' diameter), provide useful data.
1E~1743.1--2843 was within the SIS field of view only in the
observation performed on September 19, 1995. These data were analyzed to
search for the presence of spectral features, exploiting the
superior energy resolution of the CCD based detectors
(FWHM $\sim$2\% at 6 keV).

\begin{figure}[htb]
\mbox{}
\vspace{7.5cm}
\includegraphics{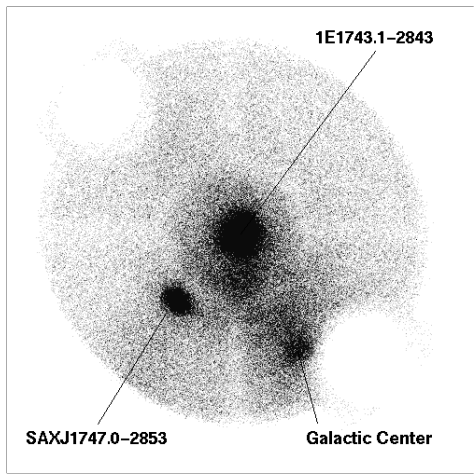}
\caption[]{$BeppoSAX$ image (2-10 keV) of the region of 1E1743.1-2843
  obtained with the MECS instrument. North is to the top, East to the 
  left. The Galactic plane runs from NE to SW.
  Note that the logarithmic color scale has been saturated to show more clearly 
  the diffuse emission that extends along the Galactic plane, between the
  SgrB2 molecular cloud (located $\sim$30  arcmin NE of 1E1743.1--2843) and
  the SgrA complex (located at the Galactic Center). 
  }
\label{MECSIMAGE}
\end{figure}

The $BeppoSAX$ image centered on 1E~1743.1--2843, obtained with the  MECS
in April 1998, is shown in Fig. \ref{MECSIMAGE}.
It is evident that 1E~1743.1--2843 lies in a region of diffuse emission,
associated both with the molecular cloud SgrB2
and with SgrA (Koyama et al. 1996; Sidoli et al. in preparation).
The fainter source $\sim$13' to the SE of 1E~1743.1--2843 is SAX~J1747.0--2853,
a transient low mass X--ray binary (Sidoli et al. 1998b).   \\
The coordinates of 1E~1743.1--2843 derived with the MECS
are $R.A.=17h~46m~19s$, $Dec.=-28\deg~43'~24''$
(J2000)  with an uncertainty of $\sim$1'. These are fully
consistent with those determined with the Einstein Observatory
(Watson et al. 1981).


\begin{table*}[htbp]
\begin{center}
  \caption{Summary of the ASCA and BeppoSAX observations of 1E~1743.1--2843.}
 
    \begin{tabular}[c]{ccccccc}
\hline
Obs.n. & Start date & Exposure & Count rate$^a$ & Off-axis   & 2-10 keV Flux${^b}$ \\
       &            & time (ks)  & (counts s$^{-1}$) & angle (') & (10$^{-10}$ erg cm$^{-2}$ s$^{-1}$)\\
\hline
A1& 30 Sep 1993 & 18.2  & 0.30  & 16 & 2.5 \\
A2& 01 Oct 1993 & 18.5  & 0.40  & 15 & 2.6 \\
A3& 01 Oct 1993 & 17.5  & 0.46  & 16 & 2.5 \\
A4& 02 Oct 1993 & 19.5  & 0.71  & 17 & 2.6 \\
A5& 15 Sep 1994 & 76.5  & 0.34  & 15 & 2.9 \\
A6& 22 Sep 1994 & 55.2  & 0.40  & 18 & 2.9 \\
A7& 24 Sep 1994 & 18.7  & 0.48  & 18 & 3.6 \\
A8& 19 Sep 1995 & 64.4  & 0.54  & 6  & 1.1 \\ 
S1& 05 Apr 1997 & 47.7  & 0.13   & 23 & 1.9 \\
S2& 24 Aug 1997 & 99.6  & 0.14   & 21 & 2.0 \\
S3& 03 Sep 1997 & 50.7  & 0.08   & 19 & 1.1 \\
S4& 13 Apr 1998 & 70.4  & 0.38   & 0  & 1.6 \\
S5& 26 Aug 1998 & 78.6  & 0.24   & 11 & 1.9 \\
\end{tabular}
\end{center}
\begin{small}
  $^{a}${In GIS2 or MECS3, not corrected for vignetting.}\\
  $^{b}${Unabsorbed flux of blackbody best fit model, corrected for vignetting.}\\
\end{small}
\end{table*}

\subsection{Spectral Analysis}

To study the source spectrum we concentrated mainly on the 
$BeppoSAX$ observation pointed on 1E~1743.1--2843 (S4),
and on the $ASCA$ one with SIS data (A8), in which we investigated
the possible presence of emission lines.

To minimize the contamination from the diffuse
emission we extracted the MECS2 and MECS3   counts
within a small radius of 2' from the source position.
The spectra were then rebinned in order to have at
least 20 counts per bin. The background spectrum
was estimated from an annular region surrounding the source.
In fact the standard MECS background spectra
obtained from blank field observations
underestimate the actual background present in the galactic center region.
As can be seen in Fig.~\ref{MECSIMAGE}, 1E~1743.1--2843 lies in a region where the
background due to the diffuse emission increases toward the
direction of the galactic center. Our choice to use a
circular corona is an effective way to estimate the likely
background value at the source position.
We verified that our main conclusions reported below do not vary
significantly by extracting the background only from the northern or
the southern half of the corona
(i.e. respectively by slightly underestimating or overestimating
the contribution of the diffuse emission).

A fit with an absorbed power law 
(photon index $\alpha \sim$2.2,
absorbing column density N$_H$~$\sim2\times 10^{23}$ cm$^{-2}$)
was unacceptable, yielding a reduced  $\chi^2$=1.49 (for 178 d.o.f.).
A  thermal bremsstrahlung  
gave a slightly better fit   ($\chi^2$=1.24)
with temperature kT$\sim$9 keV and N$_H$~$\sim2\times 10^{23}$ cm$^{-2}$.
The observed  2--10 keV flux was  $\sim1.0\times 10^{-10}$ erg cm$^{-2}$
s$^{-1}$, 
for both the power law and bremsstrahlung cases.
A good description of the  1E~1743.1--2843 spectrum
could be obtained with a blackbody model ($\chi^2$=1.01  for 178 d.o.f.).
The best fit parameters are kT=1.8 keV,
N$_H$=1.3$\times$ 10$^{23}$ cm$^{-2}$ and
flux   1.65$\times$10$^{-10}$ erg cm$^{-2}$
s$^{-1}$  (2-10 keV, corrected for absorption).
Taking  into account the uncertainties resulting
from the background subtraction, these parameters vary in the range
kT$\sim$1.7--1.9 keV, N$_{H}\sim$(1.2--1.4)$\times$10$^{23}$ cm$^{-2}$.

\begin{figure}[htb]
\mbox{}
\vspace{7.5cm}
\includegraphics{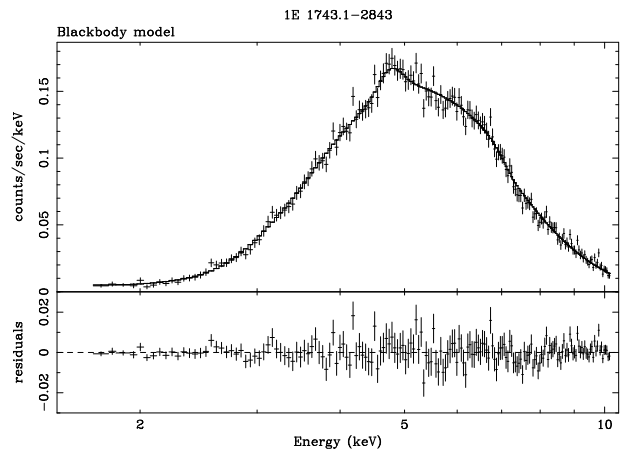}
\caption[]{Best fit spectrum of 1E~1743.1--2843
obtained with the MECS instrument.}
\label{MECSSPECTRUM}
\end{figure}

\begin{figure}[htb]
\mbox{}
\vspace{7.5cm}
\includegraphics{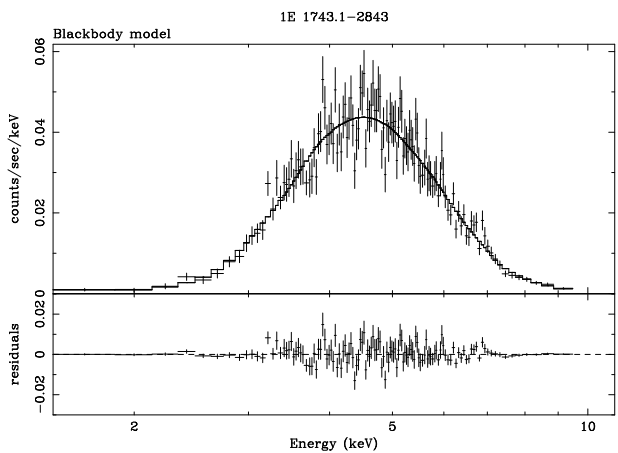}
\caption[]{Best fit spectrum of 1E~1743.1--2843 
obtained with the ASCA SIS0 instrument. The results agree with the
ones from the MECS, and are in favour of a thermal nature of the
source spectrum.}
\label{ASCASPECTRUM}
\end{figure}


The other four  $BeppoSAX$ observations, as well as the $ASCA$ GIS data
provided spectra of lower quality, due to the off-axis position of
1E~1743.1--2843 and to the lower counting statistics.
Their analysis gave results consistent with those of
obs. S4, without evidence for spectral variations.
The fluxes reported in Table 1 were therefore computed assuming for
all the observations the best fit blackbody model.

During the September 1995 $ASCA$ observation (A8), 1E1743.1-2843 was only 6' off
axis and therefore within the field of view of all $ASCA$ instruments.
The two SIS were working in 4 CCD mode with a time resolution of 16
seconds: the  data were filtered with standard criteria,
including the
rejection of hot and flickering pixels and the event grade (0,2,3 and 4)
selection.
We  extracted the source spectra from circular regions of
radius 2.5', slightly smaller than usual, in order to extract source events
from the same CCD  chip (S0C2 and S1C0) and to reduce the
background contribution.
The background spectra were   extracted from source free regions  of the
instruments during the same observation.
This procedure yielded more than 8000 counts for SIS0 and few less for
SIS1, with a background contribution of $\sim$12\% of the total.
To improve the signal to noise ratio of our spectra we grouped the  512 original
channels into only 128 bins.
In the fits we   ignored the
channels below 2 keV, where the background intensity was comparable to
that of the source itself.
After having created the appropriate response matrices (using SISRMG
v1.1 and ASCAARF v2.72), we fitted the spectra with the three single
component 
models used for the $BeppoSAX$ data analysis.
Again, the results
indicate a   thermal nature of the spectra (Fig. \ref{ASCASPECTRUM}): the   power law
fit ($\alpha \sim$ 2.2, $N_{H}\sim$2.3$\times 10^{23}$ cm$^{-2}$)
and the thermal bremsstrahlung fit
(kT $\sim$ 8.3, $N_{H}\sim$ 2.1$\times 10^{23}$ cm$^{-2}$) yielded  higher
$\chi^2$ values than the blackbody model, whose best fit parameters
(kT $\sim$ 1.65, $N_{H}\sim$ 1.5$\times 10^{23}$ cm$^{-2}$)
are similar to those obtained with the MECS,
except for the normalization.
In fact, the 
SIS data (September 1995) yielded a flux about a factor two lower than
that observed with $BeppoSAX$ in April 1998.\\
Taking advantage of the better energy resolution available with the SIS,
we searched for the presence of iron line features. We obtained
negative results: though the limited statistics does not allow to
exclude the  presence
of a broad line, we could put an upper limit of $\sim$100 eV on the
equivalent 
width of a line narrower than the SIS resolution in the 6.4--7 keV energy
range.

\subsection{Timing Analysis}

\begin{figure}[htb]
\mbox{}
\vspace{7.5cm}
\includegraphics{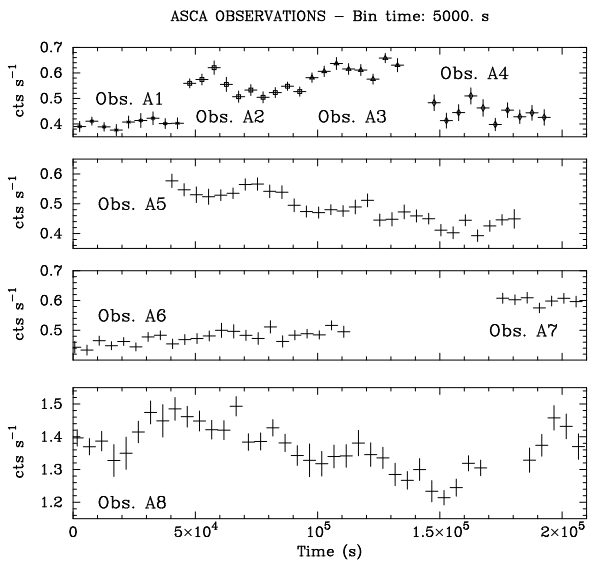}
\caption[]{Light curves of 1E1743.1-2843 (background subtracted) during the $ASCA$ observations.}
\label{ASCALC}
\end{figure}

\begin{figure}[htb]
\mbox{}
\vspace{7.5cm}
\includegraphics{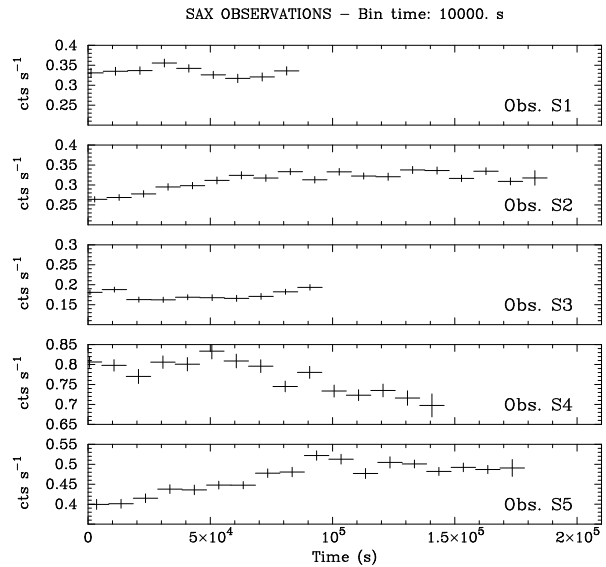}
\caption[]{Light curves  (background subtracted) during the $BeppoSAX$ observations.}
\label{SAXLC}
\end{figure}

The background subtracted light curves of 1E~1743.1--2843 obtained during
all the $ASCA$ and  $BeppoSAX$   observations are presented in Fig.~\ref{ASCALC} and
 Fig.~\ref{SAXLC}. Some variability on a time scale of hours  is clearly visible
during observations A5, A8 and S4.
Note that the count rates of Fig.~\ref{ASCALC} and \ref{SAXLC} are not corrected for vignetting
effects resulting from the different off-axis angles. This causes most of
the apparent variability between the different observations visible in the
figure, but nevertheless the source does vary considerably. As it is shown in
Table 1, the averaged and vignetting corrected fluxes of all the
observations are not consistent with a constant value, but instead
considerable 
variations up to a factor of 2 occur between different observations.


\begin{table*}
\begin{center}
\caption{Upper limits on the pulsed fraction for
different ranges of trial periods}
\begin{tabular}{lcccccc}
\hline
Observation  &  \multicolumn{6}{c}{Upper limits (99\% c.l.)} \\
     &20000\,s  & 10000\,s & 10$^3$--10$^2$\,s & 10--2\,s 
     & 1\,s--200\,ms & 100--10\,ms \\
\cline{2-7}
     & \multicolumn{6}{c}{(\%)} \\      
\hline
A1--A4  & 42 & 21 & 11    & 11     & 11 & -- \\
A5      & 55 & 30 & 10    & 10     & 10 & -- \\
A6--A7  & -- & 90 & 27--13& 15     & 15 & -- \\
A8      & 40 & 17 & 9     & 9      &  9 & -- \\
S1      & 80 & 19 & 9 & 9 & 11--14 & 14--20\\
S2      & 43 & 14 & 7 & 8 &      9 & 15 \\
S3      & 93 & 18 & 13& 13& 16--28 & 30--40 \\
S4      & 26 &  7 & 5 & 5 & 8--13  & 14--30 \\
S5      & 36 & 12 & 7 & 8 & 14     & 15--30 \\
\hline
\end{tabular}
\end{center}
\end{table*}

For each observation and detector,
we searched for periodic modulations in the X--ray flux of the source
using a Fourier analysis.
In order to increase the search sensitivity we merged the data from contiguous
observations: A1, A2, A3 and A4 as well as A6 and A7. For each data set,
after converting the arrival times  to the Solar System barycenter,
we accumulated light curves binned at 5\,ms   and calculated a single
power spectrum, correcting for the background, over the
whole observation following the method outlined by Israel \& Stella (1996).
No significant periodicity was found in any of the power spectra at a
99\% confidence level. 
In Table 2 we report the corresponding upper limits on the source
pulsed fraction for selected values of the period.

We also looked for the presence of X-ray bursts, by examining the
light curves binned at time resolutions of a few seconds, but no bursts were
found.
Though it is difficult to derive quantitative upper limits for such a kind
of analysis, we note that a typical Type~I X--ray burst with a peak luminosity
close to the Eddington limit would have produced in 1E~1743.1--2843
 an easily detectable count rate increase of a factor of a  few hundreds.

\section{Discussion}

With the lack of an optical identification, which is hampered by the high
interstellar absorption and the relatively large error region
(ROSAT did not detect this source due to the high column density,
see \cite{1994A&A...290L..29P}
), we must discuss the  nature of 1E~1743.1--2843 only on the basis
of its X--ray properties.

The first important thing to consider is the  high column density derived
from the spectral fittings. 
We notice that the X-ray source(s) at the position of the galactic center,
at an angular distance of only $\sim$16' from 1E~1743.1--2843,  
has N$_{H}\sim$7$\times$10$^{22}$ cm$^{-2}$ (Sidoli et al. 1998c).
Also the nearby  ($\sim$13') LMXRB SAX~J1747.0--2853 has a value of
N$_{H}\sim$8$\times$10$^{22}$ cm$^{-2}$. A likely distance of
the order of 10 kpc is indicated
by its persistent flux as well as by the peak luminosity during a Type~I burst
(Sidoli et al. 1998b).
It is therefore very likely that also 1E~1743.1--2843 is at a distance
of at least several kiloparsecs, and is possibly close to (or even beyond) the 
galactic center.
The resulting luminosity of $\sim$2$\times$10$^{36}$$\times$d$_{10\rm{kpc}}^2$
erg s$^{-1}$
immediately rules out models involving coronal or wind emission from
normal stars, and strongly favours the presence of an accreting
compact object.
Kawai et al. (1988) suggested  that 1E~1743.1--2843 could be a massive
binary system similar to Vela X-1 or GX301-2, and that part of the
measured absorption could be intrinsic. Instead, we believe  that the
absence of coherent pulsations (and/or eclipses) and the relatively soft spectrum
are in favour of a  low mass X-ray binary.
The observed variability is also compatible with such an interpretation.
We note in passing the striking similarity of the variability
pattern  observed in
September 1994 and 1995 and in April 1998
(see the light curves A5 and A8 of Fig.~\ref{ASCALC} and S5 of Fig.~\ref{SAXLC}).
 Unfortunately, the available data are too
sparse to establish a possible orbital periodicity as the cause of
the observed variability pattern.\\

Most LMXRB containing neutron stars are characterized by the
occurrence of Type~I X--ray bursts.
Up to now no bursts have been detected from 1E~1743.1--2843
despite it being active and repeatedly observed for over 20 years.
The $BeppoSAX$ and $ASCA$ observations reported here provide a total
effective exposure of 656 ks, in which no bursts have been detected
from  1E~1743.1--2843. Many other observations of this region of sky
obtained in recent years with imaging detectors failed
to detect bursts from this source
(for instance the ART-P instrument collected 820 ks of data,
Pavlinsky et al 1994).
If 1E~1743.1--2843 is indeed a LMXRB, the lack of bursts is
remarkable also because the most likely luminosity of this source
(10$^{36}$--10$^{37}$ erg s$^{-1}$)
is in the typical range of X--ray bursters.
The lack of bursting activity in LMXRB of higher luminosity
is generally ascribed  to the stable burning of H and He
which occurs   at luminosities close to the Eddington limit
(Fujimoto et al. 1981). Such a high luminosity would require that
1E~1743.1--2843 is at a distance greater than several tens of kiloparsecs.
%

Finally,  we cannot exclude that  1E~1743.1--2843 is an extragalactic
source seen through the Galactic plane,
which is similar to the case of GRS~1734--292 (\cite{1998A&A...330...72M}). 
An  examination of the maps of the NVSS, NRAO VLA Sky Survey (\cite{1998AJ....115.1693C}), did not reveal any possible radio
counterpart at the position of  1E~1743.1--2843 with flux
at 20 cm greater than  $\sim$50 mJy. However, considering the wide
range of X--ray to radio flux ratios observed in Active Galactic Nuclei,
the lack of a radio counterpart similar to that observed for GRS~1734--292
does not exclude the possibility of an extragalactic origin for 1E~1743.1--2843.

\begin{acknowledgements}
This research has made use of $ASCA$ data obtained through the High Energy
Astrophysics Science Archive Research Center Online Service, provided
by the NASA/Goddard Space Flight Center.
\end{acknowledgements}


\begin{thebibliography}{}

\bibitem{1}
Boella G., Chiappetti L., Conti G., et al., 1997, A\&AS 122, 327

\bibitem[Condon et al. 1998]{1998AJ....115.1693C} Condon, J. J., Cotton, W. 
D., Greisen, E. W., Yin, Q. F., Perley, R. A., Taylor, G. B. \& Broderick, 
J. J. 1998, \aj 115 1693 

\bibitem{3}
Fujimoto M.Y., Hanawa T. and Miyaji S. 1981, ApJ 246, 267

\bibitem{4}
Israel, G.L. \& Stella, L., 1996, ApJ, 468, 369


\bibitem{6}
Kawai N. et al., 1988, ApJ 330, 130

\bibitem{7}
Koyama K. et al. 1996, PASJ 48, 249

\bibitem{8}
Lu F.J. et al. 1996, A\&ASS 115, 395


\bibitem[Marti et al. 1998]{1998A&A...330...72M}
 Marti, J., Mirabel, I. F., Chaty, S. \& Rodriguez, L. F. 1998, \aa 330 72 





\bibitem{11}
Pavlinsky M.N., Grebenev S.A., \& Sunyaev R.A. 1994, ApJ 425, 110

\bibitem[Predehl \& Truemper 1994]{1994A&A...290L..29P} Predehl, P. \& 
Truemper, J. 1994, \aa 290 L29 


\bibitem{13}
Sidoli L. et al., 1998a, in "The Active X-ray Sky - Results from $BeppoSAX$
and RXTE", Nuclear Physics B (ProcSuppl.) 69/1-3, 88.

\bibitem{14}
Sidoli L., Mereghetti S., Israel G.L., Cusumano G.,
Chiappetti L., \& Treves A., 1998b, A\&A 336, L81.

\bibitem{15}
Sidoli L. et al., 1998c, in Proc. of "3rd INTEGRAL Workshop", Taormina
Sept. 1998, in press

\bibitem[Skinner et al. 1987]{1987Natur.330..544S} Skinner, G. K., 
Willmore, A. P., Eyles, C. J., Bertram, D. \& Church, M. J. 1987, \nat 
330 544 


\bibitem{17}
Sunyaev R.A. et al. 1991, A\&A 247, L29

\bibitem{18}
Tanaka Y., Inoue H. \& Holt S.S. ,1994 , PASJ 46, L37

\bibitem{19}
Watson M.G. et al. 1981, ApJ 250,142





\end{thebibliography}
\end{document}